\documentclass[conference]{IEEEtran}
\IEEEoverridecommandlockouts
\usepackage{cite}
\usepackage{amsmath,amssymb,amsfonts}
\usepackage{algorithmic}
\usepackage{tabularray}
\usepackage{graphicx}
\graphicspath{{figures/}}  
\usepackage{textcomp}
\usepackage{xcolor}
\def\BibTeX{{\rm B\kern-.05em{\sc i\kern-.025em b}\kern-.08em
    T\kern-.1667em\lower.7ex\hbox{E}\kern-.125emX}}

\makeatletter
\let\NAT@parse\undefined
\makeatother
\usepackage{hyperref}
\usepackage{booktabs}  

\begin{document}

\title{Complex Electromagnetic Space Combat System-of-systems Modeling and Key Node Identification Method

\thanks{This work was supported by the National Natural Science Foundation of China under Grants 62201607.}
}

\author{\IEEEauthorblockN{Xiao Liu}
\IEEEauthorblockA{
\textit{Academy of Military Science}\\
Beijing, China \\
liuxiao\_0817@163.com}
\and
\IEEEauthorblockN{Sudan Han}
\IEEEauthorblockA{
\textit{Academy of Military Science}\\
Beijing, China \\
xiaoxiaosu0626@163.com}
\and
\IEEEauthorblockN{Jinlin Peng*}
\IEEEauthorblockA{
\textit{Academy of Military Science}\\
Beijing, China \\
peng\_jinlin@126.com}
}

\maketitle

\begin{abstract}
With the application of advanced science and technology in the military field, modern warfare has developed into a confrontation between systems. The combat system-of-systems (CSoS) has numerous nodes, multiple attributes and complex interactions, and its research and analysis are facing great difficulties. Electromagnetic space is an important dimension of modern warfare. Modeling and analyzing the CSoS from this perspective is of great significance to studying modern warfare and can provide a reference for the research of electromagnetic warfare. In this study, the types of nodes and relationships in the complex electromagnetic space of CSoS are first divided, the important attributes of the combat nodes are extracted, and the relationship weights are normalized to establish a networked model. On this basis, the calculation method of CSoS combat effectiveness based on the combat cycle is proposed, and then the identification and sorting of key nodes can be realized by the node deletion method. Finally, by constructing an instance of aircraft carrier fleet confrontation, the feasibility of this method has been verified, and the experimental results have been compared with classical algorithms to demonstrate the advanced nature of this method.
\end{abstract}


\section{Introduction}
With the continuous development of advanced technology in the military field, modern warfare has evolved from weapon-to-weapon and individual-to-individual confrontation to the overall confrontation between combat systems of systems (CSoS)~\cite{lijichao-1}. However, the number of nodes in the combat system is huge, the performance attributes are multi-dimensional, and the correlation is complex, thus the behavioral interaction between the CSoS can no longer be measured by the simple superposition of multiple node pairs. The overall combat effectiveness has a significant emergence character~\cite{huxiaofeng}, which means the increase or decrease of a certain key point or even the change of attributes may affect the victory or defeat of the entire war. Finding out such key points in the CSoS will help the defender strengthen the defense of the weak links and the attacker carry out more cost-effective attacks. This problem is of great value to both the offensive and defensive sides in the war, so it has always been a hot spot in the research of CSoS.

At present, the research on finding the key points of CSoS is extensive. Some researchers use the method of network science to study the CSoS~\cite{tanyuejin-1} and use the structural parameters of the complex network of the CSoS to find the key points, such as the parameter of the node (degree, betweenness, closeness centrality, PageRank~\cite{pagerank}, etc.) , or the degree of change of the overall structural parameters of the network (such as aggregation coefficient\cite{watts-1}, average path length~\cite{watts-2}, directed natural connectivity~\cite{lijichao-2}) after deleting, dividing or shrinking the network~\cite{tanyuejin-2}, to determine the importance of the node. However, this method does not take into account the heterogeneity of nodes in the combat network and the practical significance of the real node association relationship, and the interpretability of the algorithm is weak.
Some researchers use the OODA (Observation, Orientation, Decision, Action) cycle theory to analyze the CSoS~\cite{tanyuejin-1}. References \cite{zhangchunhua,zhaodanling} quantitatively describe the relationship between equipment nodes with the help of task satisfaction, then establish a network model, and find the key nodes of the equipment system according to the number of combat loops and capability satisfaction. Reference ~\cite{wangyaozu} proposed a network state transformation method based on Monte Carlo sampling, which realized the network structure modeling of transforming edge weights into edge existence, and evaluated the importance of nodes according to the comprehensive ability of the system formed by the killing chain. Another search algorithm for combat cycles based on an adjacency matrix and an effectiveness evaluation method for combat cycles based on improved information entropy has been proposed~\cite{luochengkun}. These methods are popular approaches for analyzing CSoS, but only applicable in conventional domains.

In the context of highly information-driven modern warfare, network-centric information systems have emerged as the fundamental architecture for contemporary CSoS. The research mentioned above is mostly concentrated in physical and logical space, and the electromagnetic space in which these equipment are mainly located has become an important dimension of modern warfare. Few scholars have studied the CSoS from this perspective. At present, the basic research on radar detection, communication, and electromagnetic interference in complex electromagnetic space is relatively mature. References~\cite{donghaohao,houdaoqi} theoretically modeled the detection probability of radar target detection. References~\cite{liwei,liuqi-1,liuqi-2} theoretically analyzed and simulated the communication and anti-jamming performance of the Link16 data link in the active service of the U.S army. References~\cite{jinjing,kao} modeled and simulated the anti-jamming effect of the Link16/JTIDS communication system under different interference signals in different channel environments. The above research is aimed at the analysis and calculation of specific technology or combat links, but the analysis of the combat process corresponding to the CSoS is relatively isolated. In addition, the modeling of specific technologies may be too complex. When considering the establishment of the CSoS model, because of its many types of nodes and complex correlation, it is necessary to consider the balance between model accuracy and computational complexity, and a more general and appropriate mathematical model should be established.

Given the above shortcomings, this paper employs the concept of networked modeling and analysis to propose a quantitative modeling method for CSoS within complex electromagnetic space. Building on this foundation, the paper proposed a method for calculating the combat capability of CSoS based on the combat cycle's capability and a key node identification method, followed by verification through examples.

\section{Networked modeling}

 The modern combat cycle theory holds that the combat process is a cycle process composed of Observation-Orientation-Dicision-Action (OODA), i.e. the reconnaissance node discovers the enemy target and transmits the information to the decision-making node, the decision-making node sends commands to the attack node after analyzing and decision-making, and the attack node attacks the enemy target node after receiving the attack command. Thus, this chapter begins with an exposition on the classification of nodes and the extraction of key attributes of the nodes, followed by a discussion on the categorization and computation of the various relationships between nodes.

\subsection{Modeling of nodes}

In the complex electromagnetic space, the nodes can be divided into the following according to the types of function~\cite{cares}:

\textit{1) Detection node (S, Sensor)}. Responsible for detecting and locating enemy targets. As the most common and widely used reconnaissance and detection equipment in modern warfare, radar plays a key role in reconnaissance and intelligence systems at all levels. Therefore, we use radar as the main modeling object of reconnaissance and detection nodes.

\textit{2) Communication/decision node (C, Communication/ D, Decider)}. Equipment with communication function. In this paper, each node has a communication module and can be used as a relay to achieve multi-hop communication between long-distance nodes. If the node has command and control function, it is also a decision-making node. 

\textit{3) Interference node (I, Influencer)}. Equipment node performing electromagnetic attack. Interference methods such as suppression interference can be performed.

\textit{4) Target node (T, Target)}. A node within the target CSoS that is a candidate for our detection and attack. It can also be divided into the above three categories.

\begin{figure}[htbp]
\centerline{\includegraphics[width=0.48\textwidth]{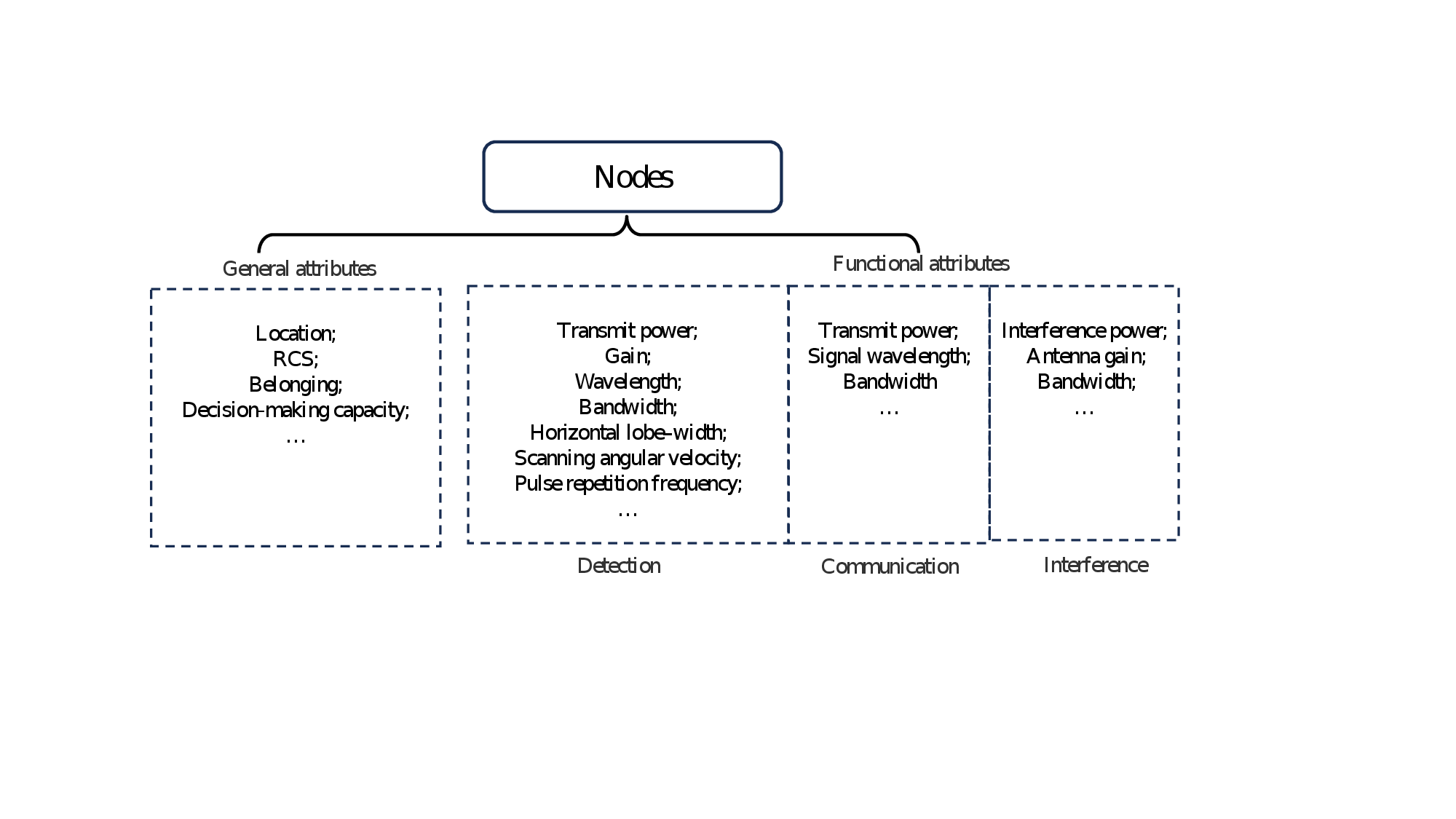}}
\caption{Attribute classification of nodes.}
\label{fig1}
\end{figure}

The attributes of the above node types can be divided into two categories, one is general attributes, and the other is functional attributes. General attributes include node position, radar cross section (RCS), etc. Functional attributes include detection function attributes, communication and decision attributes, and interference attributes. The detailed attribute classification is shown in the Fig.\ref{fig1}. Among them, a node can have and play multiple functions at the same time. On the same equipment node, there may be multiple types of equipment with the same functional type, but we take the equipment with more advanced performance as the main modeling object. The attributes of enemy targets are the same.

\subsection{Modeling of relations}

In most of the past studies, the relationship between nodes has only two states of 0 and 1 (existence and non-existence), and Zhao et al. used task satisfaction to calculate the weight of the edge, i.e. according to the attributes of the two nodes~\cite{zhaodanling}. The ability value is compared with the demand value, and the weight of the edge is calculated between 0 and 1. This method is convenient for calculation, but it needs a preset task ability demand value system, which needs to be objective and reasonable as far as possible. In the modeling of relationships, this study characterizes the weights of edges as the effectiveness or probability of the relationship between nodes, which makes the meaning of the weights of edges more intuitive and practical.

To establish a networked model of CSoS, according to the above node types, the following types of relationships are quantitatively analyzed.

\textit{T-S}: detection discovery relationship. The intuitive meaning of capability is the probability that the enemy target node is discovered by the detection node. When the false alarm rate is ${{10}^{-6}}$, the detection probability of the radar to the target can be expressed as (1).
\begin{align}
  {{P}_{d}}=1-\text{}\Phi\text{}\left(\frac{4.75-\sqrt{n}{{S}_{N}}}{\sqrt{1+2{{S}_{N}}}}\right),
\end{align}
where $\Phi(x)=\frac{1}{\sqrt{2\pi}}\int_{-\infty}^{x}\exp\left(\frac{-t^{2}}{2}\right)\mathrm{d}t$, $n$ denotes the number of echoes received by the radar in each scan, $n=\frac{{{\theta }_{0.5}}\cdot {{f}_{r}}}{\Omega }$, ${{S}_{N}}=\frac{{{P}_{t}}G_{t}^{2}{{\lambda }^{2}}\sigma }{{{(4\pi )}^{3}}R_{t}^{4}\text{ }\cdot \text{ }K{{T}_{0}}\text{ }\Delta \text{ }{{f}_{r}}{{F}_{n}}L}$, ${{\theta}_{0.5}}$ denotes the horizontal lobe width~\cite{shaoguopei}.

\textit{S-D, D-D, D-I}: communication / command relations. The relationship includes the following four cases, the reconnaissance and detection nodes transmit the target information to the decision-making nodes, decision-making nodes transmit information between decision-making nodes, relay nodes receive and forward information, and decision-making nodes transmit the attack information to the interference nodes. The intuitive meaning of capability is the quality and connectivity of communication.

When the communication receiver is without interference, the signal-to-noise ratio (SNR) of the communication signal can be expressed as:
\begin{align}
SNR=\frac{{{P}_{r}}}{\sigma _{b}^{2}},
\end{align}
where ${{P}_{r}}=\frac{{{P}_{t}}{{G}_{t}}{{G}_{r}}{{\lambda }^{2}}}{{{(4\pi )}^{2}}{{d}^{2}}L}$, $\sigma _{b}^{2}$ denotes the white noise power.  In military communication, in order to enhance the anti-interference ability, the equipment is often integrated with various technologies, such as CRC code, spread spectrum, and frequency hopping of Link16 system~\cite{liwei}, which makes the communication rate calculation complicated. And in real-world combat, the connectivity and reliability of communication is more important than the transmission speed. The bit error rate (BER) is a key indicator to measure the quality of communication, which can be used to measure communication capability. To avoid complex and inefficient calculation, the relationship between BER and SNR can be determined by simulation or experiment, and then the function model of communication efficiency and BER can be established, whose value is between 0 and 1. Specific simulation or experimental methods for different communication systems and equipment are not discussed here.

\begin{figure}[htbp]
\begin{minipage}{0.49\linewidth}
\vspace{3pt}
\centerline{\includegraphics[width=\textwidth]{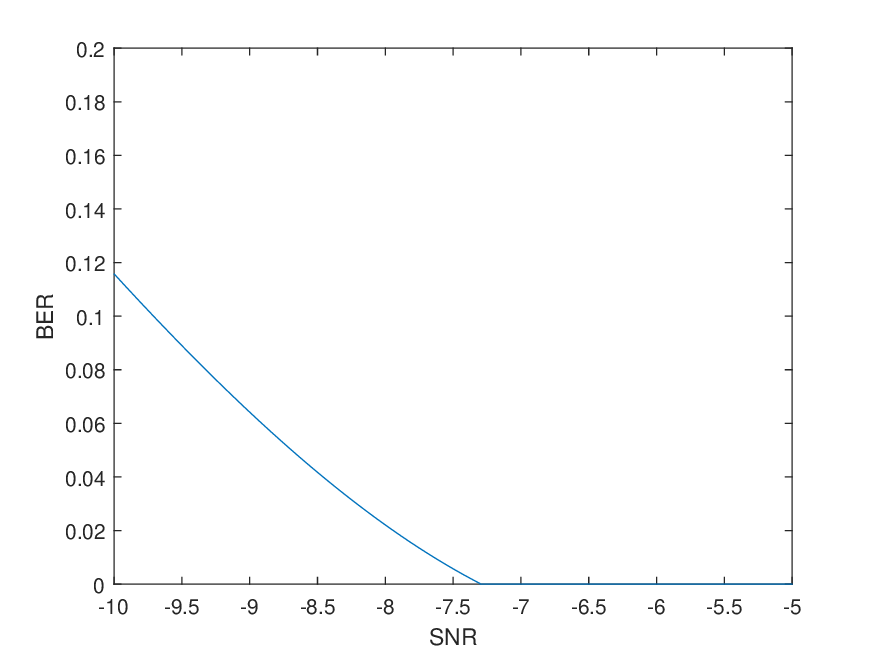}}
\centerline{(a)}
\end{minipage}
\begin{minipage}{0.49\linewidth}
\vspace{3pt}
\centerline{\includegraphics[width=\textwidth]{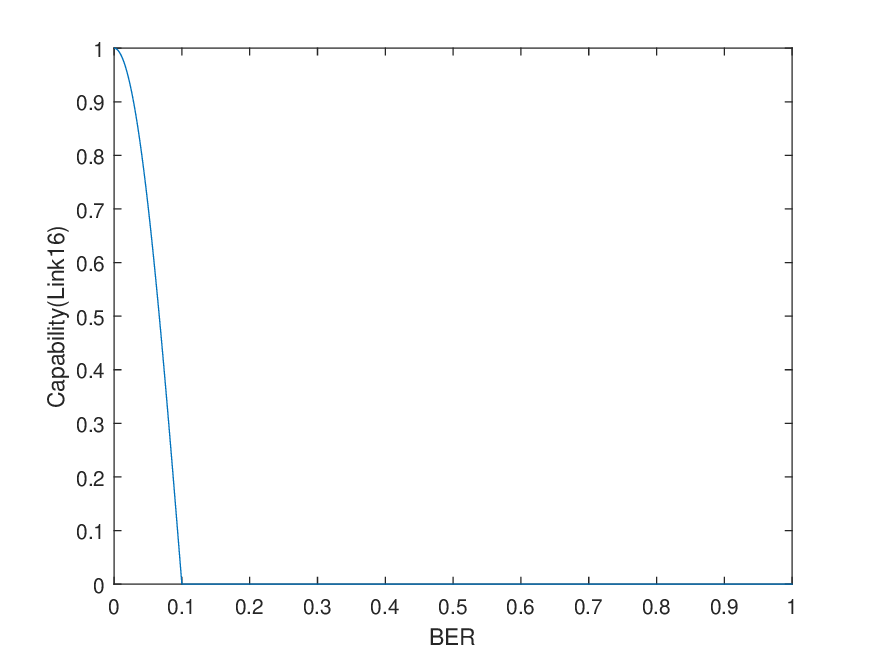}}
\centerline{(b)}
\end{minipage}
\caption{(a) Fitting function images of SNR and BER;(b) Function image of BER and communication efficiency}
\label{fig2}
\end{figure}

Taking the Link16 data link as an example, the efficiency value can be calculated by using the simulation experiment data which is closer to the real situation. For the convenience of calculation, as is shown in Fig.\ref{fig2}(a), the relationship between bit error rate and interference-to-signal ratio under multi-narrowband noise interference can be approximated by an apolynomial function~\cite{liwei}~\cite{liuqi-1}.

Generally, when the bit error rate is greater than $0.1$, the bit error ratio has exceeded the error correction capability of the system, We believe that the communication relationship between the two nodes cannot be established, the communication efficiency is $0$. When the bit error rate is less than $0.1$, the cosine function is used as the normalization function, and the relationship between the communication relationship efficiency and the bit error rate is defined as (3).The function image is shown in Fig.\ref{fig2}(b).
\begin{align}
Cap_{comm}=\cos (5\pi \times BER).
\end{align}
When the performance data of the communication system is easy to calculate, the communication capability can be obtained by direct simulation or calculation. However, most of the time in real life, such performance data is difficult to obtain through simple calculation and simulation, and the polynomial function fitting method is more operable.

\textit{I-T}: Interference relationship. The core of electromagnetic interference effectiveness evaluation is to measure the change in the ability of the target node after interference. The interference signal transmitted by the target node is usually not affected by noise, so it is assumed that our interference does not affect the enemy's interference efficiency, i.e. the interference node can only affect the enemy's communication and detection capabilities.Assuming that the performance data of the interfered party is transparent and known.

\begin{figure}[htbp]
\centerline{\includegraphics[width=0.35\textwidth]{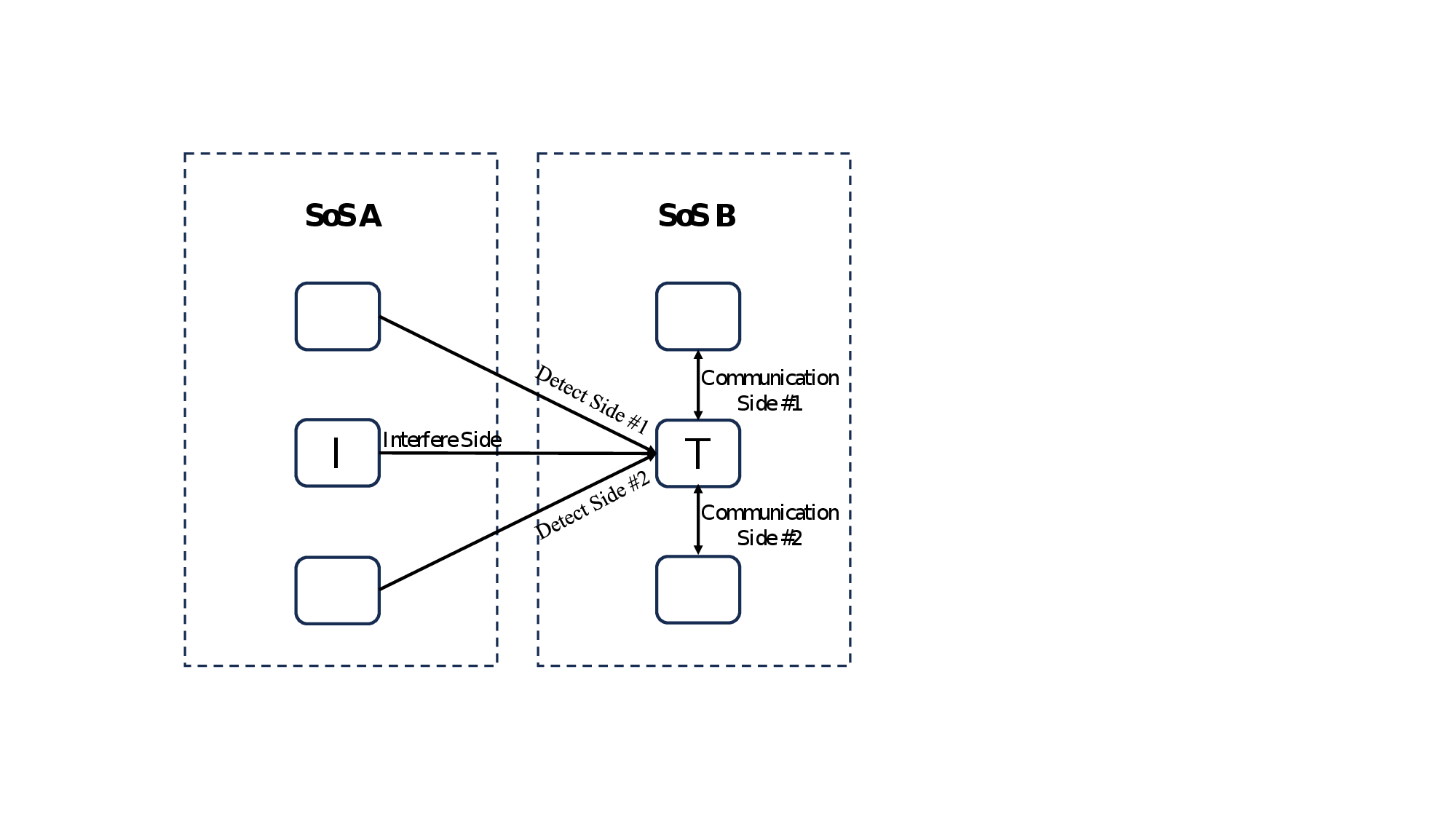}}
\caption{Example of an interference relation.}
\label{fig3}
\end{figure}

When the target node only has a single function, i.e. it is only on one edge of the detection or communication relationship, the change ratio of the capability of the edge can be directly used to measure the interference capability. However, as is shown in Fig.\ref{fig3}, in a complex CSoS, the target node is likely to be in multiple communication and detection relationship edges at the same time. When calculating the interference capability, it is necessary to consider the effectiveness change of each edge of the target node before and after interference.
Firstly, the efficiency calculation methods of detection and communication relationship under interference should be discussed respectively. The calculation of the interference efficiency of the communication relationship is based on the signal-to-interference ratio. The interference signal power received by the target node is:
\begin{align}
{{P}_{jr}}=\frac{{{P}_{j}}{{G}_{jr}}{{G}_{rj}}{{\lambda }^{2}}}{{{(4\pi {{d}_{j}})}^{2}}}.
\end{align}
The interference power is substituted into (2) to obtain the signal-to-interference ratio. Based on the calculation method of the signal-to-interference ratio and the communication relationship model above, the performance of the communication relationship of the target node after interference can be obtained. The calculation of the detection relationship capability under interference is similar to the methods of calculation of the communication relationship. According to (1), the detection probability under interference can be calculated. 

On this basis, the change rate of the relationship effectiveness value is used to represent the effectiveness of our interference node on a certain side interference effect.
\begin{align}
\overset{k\to j}{\mathop{Cap_{i\to j}}}\,=\frac{Cap_{k\to j}^{'}-Cap_{k\to j}}{Cap_{k\to j}}.
\end{align}
Node $i$ is an interference node, $j$ is the target node, and $k$ is a node with an edge between $j$ in the CSoS. $Cap_{k\to j}$ and $Cap'_{k\to j}$ represent the capability value of the relationship edge from $k$ to $j$ before and after interference respectively, and $\overset{k\to j}{\mathop{Cap_{i\to j}}}$ represents the interference capability value of node $i$ on a certain relationship between node $k$ and $j$. 

In order to reduce the adverse effect of the drastic change of the edge with a lower capability value, when calculating the interference capability of the target node in multiple relationship edges, the original capability value of the relationship edge is added as the weight, and the interference capability of the interference node to the target node is expressed as the weighted average of the capability change rate of each relationship edge:
\begin{align}
Cap_{i\to j}=\underset{k=1}{\overset{N}{\mathop \sum }}\,\overset{k\to j}{\mathop{Cap_{i\to j}}}\,\frac{Caps_{k\to j}}{\mathop{\sum }_{\text{n}=1}^{N}Cap_{\text{n}\to j}},
\end{align}
where $N$ is the total number of nodes $k$ associated with $j$.

\section{Capability calculation and key node identification}

\subsection{Capability calculation of a single combat cycle}
In a combat cycle, the relationship between each party is serial, so each edge on a link needs to have non-zero combat capability at the same time, so that the whole link has combat capability. At this time, the above nodes and edges form a combat cycle. The change of the capability value of any edge contained in the combat cycle will change the overall capability of the combat cycle. The capability of a single combat cycle is obtained by multiplying the capability values of all the relationship edges contained in the cycle\cite{zhaodanling}.
In the actual combat process, the length of the combat cycle will not be infinitely long, and even if there is a too-long combat cycle, due to the low reliability and timeliness, the impact on the combat effect is also small. Therefore, the penalty factor can be added to reduce the effectiveness of the too-long combat cycle and reduce its impact on the evaluation.
\begin{equation}
Cap{_{cycle}}=\underset{e\in {{E}_{cycle}}}{\overset{{}}{\mathop \prod }}\,Cap{_{e}}\cdot P{_{e}},
\end{equation}
where $Cap{_{e}}$ denotes the capability of the edge $e$, $P{_{e}}$ denotes the punishment of the negative effect due to the increase in the length of the combat cycle.
\subsection{Capability calculation of the CSoS}
Intuitively, the more the number of combat cycles, the more ways the CSoS can accomplish a certain task, the stronger the robustness. At the same time, the higher the effectiveness of these combat cycles, the greater the probability of completing the task. 
The relationship between the vast majority of combat cycles can be regarded as parallel at the macro level. For example, if the combat mission cannot be completed, all combat cycles need to fail. Based on the above ideas, the overall combat capability of the system is set as (8).
\begin{equation}
Cap_{CSoS}=1-\underset{cycle\in C}{\overset{{}}{\mathop \prod }}\,(1-Cap_{cycle}),
\end{equation}
where $C$ is the set of all combat cycles in the CSoS. If the enemy's CSoS has multiple nodes, the weight can be preset according to their importance. The comprehensive combat capability of our CSoS against the enemy CSoS can be expressed as (9).
\begin{equation}
    Cap_{CSoS}=\underset{target\in T}{\overset{{}}{\mathop \sum }}\,{{w}_{target}}\cdot Cap_{CSoS\to target},
\end{equation}
where ${{w}_{target}}$ is the weight of the target node in the enemy CSoS in the combat task.

\subsection{Key nodes identification based on node deletion method}
Based on the original CSoS, one of the nodes is deleted once, and the importance of the node is measured by calculating the degree of ' weakening ' it brings to the overall capability of the CSoS~\cite{tanyuejin-2}.

\begin{equation}
    CRT_{X}^{SoS}=\left\{ \begin{matrix}
   0,Cap_{SoS-\{X\}}>Cap_{SoS},  \\
   \frac{Cap_{SoS}-Cap_{SoS-\{X\}}}{Cap_{SoS}}\times100\%, else,  \\
\end{matrix} \right.
\end{equation}
where $CRT_{X}^{CSoS}$ represents the level of criticality of node $X$ to CSoS, $Cap_{CSoS}$ represents the combat capability of CSoS, and $Cap_{CSoS-\{X\}}$ represents the combat capability of CSoS after node $X$ is deleted.

\section{Case Study}
\subsection{Case background}
We constructed an aircraft carrier fleet confrontation case as a research example. The aircraft carrier formation of A CSoS cruises in a certain area of the sea, including 1 cruiser and 3 destroyers around the aircraft carrier, 1 early warning aircraft is sent for reconnaissance and early warning tasks, 3 batches of 6 fighters aircraft and 1 electronic warfare aircraft are sent for long-range patrol; the B CSoS sent fighters to reconnaissance and strike the A CSoS, including 1 early warning aircraft for remote command and support, 1 batch of 1 fighter, and 1 electronic warfare aircraft to perform strike missions \cite{zenglikai}. The code, number and function of nodes are shown in Table~\ref{tab:function}. The reference source of equipment parameters is published by open source~\cite{723}.


\renewcommand\arraystretch{1.0}   
\begin{table}[htbp]
\centering
\caption{The quantity and function of equipment in CSoS}
\label{tab:function}
\resizebox{0.48\textwidth}{!}{
\begin{tabular}{cccccccc}
\toprule
\textbf{Names}   & \textbf{Stand} &\textbf{Code}& \textbf{Num} & \textbf{S} & \textbf{C} & \textbf{I}  &\textbf{D}\\ 
\midrule
Aircraft Carrier     &A     &1    &1   &\checkmark &\checkmark &\checkmark &\checkmark \\
Cruiser                     & A      & 2    & 1      & \checkmark        & \checkmark & \checkmark &   \\
Destroyer                   & A      & 3-6  & 4      & \checkmark        & \checkmark & \checkmark &   \\
Early warning aircraft      & A      & 7    & 1      & \checkmark        & \checkmark &   & \checkmark \\
Fighter                     & A      & 8-13 & 6      & \checkmark        & \checkmark &   &   \\
Electronic warfare aircraft & A      & 14   & 1      & \checkmark        & \checkmark & \checkmark &   \\
Fighter                     & B      & B:F  & 1      & \checkmark        & \checkmark &   &   \\
Early warning aircraft      & B      & B:W  & 1      & \checkmark        & \checkmark &   &   \\
Electronic warfare aircraft & B      & B:E  & 1      & \checkmark     & \checkmark & \checkmark \\
\bottomrule
\end{tabular}
}
\end{table}

\subsection{Relation computing and network modeling}
\begin{figure*}[htbp]
\centerline{\includegraphics[width=0.90\textwidth]{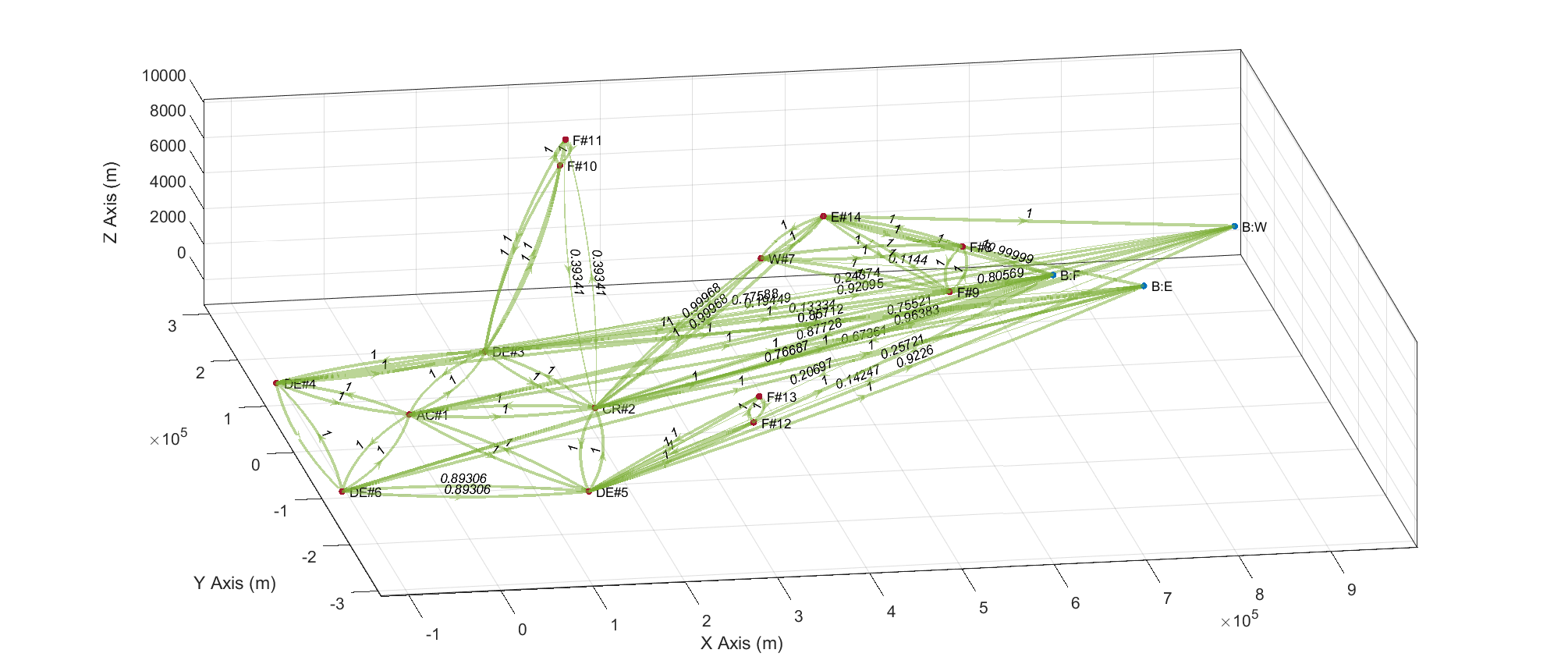}}
\caption{Schematic diagram of the networked CSoS model with only nodes and directed weighted edges.}
\label{fig:network}
\end{figure*}

Using the above methods, adjacency matrix for multiple relationships can be obtained, thus a networked CSoS model is formed, and the image representation is shown in Fig.~\ref{fig:network}.
\subsection{CSoS capability calculation and key point identification}
Taking the node weights of CSoS B as $0.3$, $0.4$, and $0.3$ respectively as an example for calculation. Since the aircraft carrier is the core equipment of the A CSoS,  the destruction of the aircraft carrier is not considered. The capability of the benchmark CSoS and the capability based on the node deletion method are shown in Table~\ref{tab:result}.
\renewcommand\arraystretch{1.0}   
\begin{table}[htbp]
\centering
\caption{Capability value of benchmark CSoS and its comparison by node deletion}
\label{tab:result}
\resizebox{0.35\textwidth}{!}{
\begin{tabular}{ccc}
\toprule
\textbf{Delete Node}   & \textbf{CSoS capability} &\textbf{Criticality}\\ 
\midrule
None & 0.9580 & ~/      \\
2    & 0.6903 & 30.59\% \\
3    & 0.9573 & 3.74\%  \\
4    & 0.9812 & 1.34\%  \\
5    & 0.9595 & 3.52\%  \\
6    & 0.9820 & 1.26\%  \\
7    & 0.9899 & 0.46\%  \\
8    & 0.9938 & 0.07\%  \\
9    & 0.9938 & 0.07\%  \\
10   & 0.9945 & 0.00\%  \\
11   & 0.9943 & 0.02\%  \\
12   & 0.9945 & 0.00\%  \\
13   & 0.9945 & 0.00\%  \\
14   & 0.9902 & 0.43\%  \\
\bottomrule
\end{tabular}
}
\end{table}

The results of CRT index can be summarized as but not limited to the following conclusions: 
The cruiser is the most critical node, and the fighters are the least critical nodes. The ships are more critical than the aircrafts; early warning aircraft and electronic warfare aircraft are more critical in aircraft nodes because of their functional characteristics. The above results are the same as the intuition and experience of experts.

\subsection{Algorithms comparison}
The following key node identification indicators were calculated respectively: The number of combat cycles based on node deletion method; the node degree centrality, closeness centrality, betweeness centrality, Pagerank (transition probability is set to $0.5$), and eigenvector centrality of the communication network. The results are as shown in Table \ref{tab:compare}.
\renewcommand\arraystretch{1.0}   
\begin{table*}[htbp]
\centering
\caption{Value and rank of CSoS nodes’ Criticality by CRT and other classic methods}
\label{tab:compare}
\resizebox{0.90\textwidth}{!}{
\begin{tabular}{cccccccccccccc}
\toprule
\multicolumn{2}{c}{CRT}&\multicolumn{2}{c}{Combat cycles}&\multicolumn{2}{c}{Degree}  &\multicolumn{2}{c}{Closeness}     &\multicolumn{2}{c}{Betweenness}      &\multicolumn{2}{c}{PageRank}
&\multicolumn{2}{c}{Eigenvector}\\
value     & rank & value & rank & value & rank & value & rank & value & rank & value  & rank & value  & rank  \\
\midrule
30.5882\% & 2    & 89    & 2    & 6     & 2    & 0.05  & 2    & 43.5  & 2    & 0.0919 & 5    & 0.1249 & 2     \\
3.7406\%  & 3    & 143   & 3    & 5     & 3    & 0.042 & 5    & 24.75 & 5    & 0.0902 & 3    & 0.1024 & 3     \\
3.5194\%  & 5    & 150   & 5    & 5     & 5    & 0.04  & 3    & 12    & 3    & 0.0863 & 2    & 0.0938 & 5     \\
1.3374\%  & 4    & 206   & 4    & 4     & 7    & 0.037 & 7    & 10    & 7    & 0.0748 & 7    & 0.0793 & 7     \\
1.2569\%  & 6    & 206   & 6    & 4     & 14   & 0.037 & 14   & 10    & 14   & 0.0748 & 14   & 0.0793 & 14    \\
0.4625\%  & 7    & 221   & 7    & 3     & 4    & 0.036 & 11   & 3.5   & 11   & 0.0653 & 8    & 0.0738 & 4     \\
0.4324\%  & 14   & 236   & 14   & 3     & 6    & 0.032 & 4    & 1.5   & 6    & 0.0653 & 9    & 0.07   & 6     \\
0.0704\%  & 8    & 264   & 8    & 3     & 8    & 0.032 & 6    & 1.25  & 4    & 0.064  & 4    & 0.054  & 8     \\
0.0704\%  & 9    & 264   & 9    & 3     & 9    & 0.029 & 12   & 0     & 8    & 0.0631 & 6    & 0.054  & 9     \\
0.0201\%  & 11   & 279   & 11   & 3     & 11   & 0.029 & 13   & 0     & 9    & 0.0623 & 11   & 0.0482 & 11    \\
0.0000\%  & 10   & 286   & 10   & 2     & 10   & 0.028 & 10   & 0     & 10   & 0.0602 & 12   & 0.0383 & 10    \\
0.0000\%  & 12   & 286   & 12   & 2     & 12   & 0.027 & 8    & 0     & 12   & 0.0602 & 13   & 0.0319 & 12    \\
0.0000\%  & 13   & 286   & 13   & 2     & 13   & 0.027 & 9    & 0     & 13   & 0.0578 & 10   & 0.0319 & 13    \\
\bottomrule
\end{tabular}
}
\end{table*}

The results of evaluating the importance of nodes by using the number of combat cycles are similar to the $CRT$ indicators, and both can be better confirmed by the empirical speculation. This shows that it is necessary to use the theory of combat cycle to study, but the method of $CRT$ effectiveness evaluation can quantify the criticality of nodes more intuitively and precisely.
The other methods have certain limitations: the methods using degree centrality, closeness centrality, betweeness centrality and eigenvector centrality are based on the communication network topology, while ignore the important role of some nodes with detection and interference capabilities (4, 6, etc.); the top few key points obtained by PageRank method are contrary to experience and have poor interpretability. In addition, the above methods will produce tied ranking results, which makes the key point recognition ineffective.

\section{Conclusion}
In response to the issue of CSoS modeling and key node identification in complex electromagnetic space, this paper proposes a comprehensive modeling and analysis method with the following advantages: 1) The modeling method is highly operable and versatile, capable of transforming dispersed military equipment into a directed weighted graph model. 2) It can precisely quantify the overall performance of CSoS and provide a ranking of node criticality based on the node deletion method. 3) Example verification shows that this method has strong interpretability, and the identification performance of key points is superior to traditional methods. 
Therefore, this paper has certain application and research value in the study of CSoS and the selection of combat targets. 

However, the current modeling method is limited to the electromagnetic domain, and future research may consider incorporating more types of relationships. The current method for identifying key nodes is static, and future study may include task factors.

\end{document}